\documentclass[prb,twocolumn,showpacs,preprintnumbers,amsmath,amssymb]{revtex4}
\usepackage{graphicx}
\usepackage{dcolumn}
\usepackage{bm}

\begin{document}

\preprint{preprint}

\title{Superfluid, Mott-Insulator, and Mass-Density-Wave
Phases \\ in the One-Dimensional Extended Bose-Hubbard Model}

\author{Ramesh V. Pai}
\email{rvpai@unigoa.ac.in}
\affiliation{
Department of Physics, Goa University, Goa 403 206, India.
}

\author{Rahul Pandit}
\email{rahul@physics.iisc.ernet.in}
\altaffiliation [Also at ]{Jawaharlal Nehru Centre for Advanced Scientific
  Research, Jakkur, Bangalore 560 064, India.
}
\affiliation{
Centre for Condensed Matter Theory, Department of Physics,
  Indian Institute of Sciences,
  Bangalore 560 012, India
}

\date{\today}

\begin{abstract}
  We use the finite-size density-matrix-renormalization-group (FSDMRG) 
  method to obtain the phase diagram of the one-dimensional 
  ($d = 1$) extended Bose-Hubbard model for density $\rho = 1$
  in the $U-V$ plane, where $U$ and $V$ are, respectively, onsite
  and nearest-neighbor interactions. The phase diagram comprises
  three phases: Superfluid (SF), Mott Insulator (MI) and Mass Density Wave
  (MDW). For small values of $U$ and $V$, we get a reentrant SF-MI-SF
phase transition. For intermediate values of interactions the SF phase is
  sandwiched between MI and MDW phases with continuous SF-MI and SF-MDW
  transitions. We show, by a detailed finite-size scaling analysis, that the
  MI-SF transition is of
  Kosterlitz-Thouless (KT) type whereas the MDW-SF transition has 
  both KT and two-dimensional-Ising characters.
  For large values of $U$ and $V$ we get a direct, first-order, MI-MDW 
transition. The MI-SF, MDW-SF and MI-MDW phase boundaries join at a 
bicritical point at ($U ,V ) = (8.5 \pm 0.05 ,4.75 \pm 0.05)$. 

\end{abstract}

\maketitle

\section{Introduction}

The study of quantum phase transitions in systems of interacting bosons is an 
exciting area with a fruitful interplay between 
theory~\cite{ma,fisher,krauth,sheshadri,tvr,amico,rvpai,
baltin,kuhner,kashurnikov}, numerical 
simulations~\cite{batrouni,nandini,wallin,zhang}, and experiments. 
A variety of experimental systems have been studied: liquid $^4$He in porous 
media like vycor or aerogel~\cite{chan}; microfabricated 
Josephson-junction arrays~\cite{chow,Mooji}; the 
disorder-driven superconductor-insulator transition in thin films of 
superconducting materials like bismuth~\cite{goldman}; flux 
lines in type-II superconductors pinned by columnar defects 
aligned with an external magnetic field~\cite{nelson}; 
and best from the point of view of comparing theory with experiments,
atoms trapped in optical-lattice potentials. In a system where the number of
atoms per site is an integer, Greiner \textit{et al.}~\cite{greiner} have 
observed a superfluid-Mott insulator transition for $^{87}$Rb atoms, trapped 
in 
a three-dimensional optical-lattice potential, by changing the strength of 
the onsite potential. Experiments in such optical lattices have several 
advantages over their condensed-matter counterparts, including precise 
knowledge 
of the underlying microscopic models~\cite{jaksch}, the possibility of
controlling parameters in the effective lattice Hamiltonians, and
the absence of disorder. The recent probable observation of a supersolid 
helium phase~\cite{kim} has given a further fillip to this area.
Even in the absence of disorder these systems can show a variety of phases 
like Superfluid (SF), Mott Insulator (MI), and Mass Density Wave (MDW) 
[or Charge Density Wave (CDW) if the bosons are charged]. The simplest model 
that can show these  phases is the extended Bose-Hubbard model whose 
Hamiltonian is
\begin{eqnarray}
  \label{eq:ebhmodel}
  {\cal H} &=&-t \sum_{<i,j>}  (a_i^\dagger a_j +h.c )
  + \frac{U}{2}\sum_i {\hat n}_i
  ({\hat n}_i-1) \nonumber \\ && 
+ V \sum_{<i,j>} {\hat n}_i {\hat n}_j.
\end{eqnarray}
The first term in Eq.~(\ref{eq:ebhmodel}) represents the kinetic energy 
associated with the hopping of bosons from site $i$ to its nearest-neighbor 
site $j$ with amplitude $t$;  $a_i^\dagger$ ($a_i$) is the boson creation 
(annihilation) operator at site $i$ and ${\hat n}_i=a_i^\dagger a_i$ is the 
associated number operator; onsite $U$ and nearest-neighbor $V$ interactions 
are represented, respectively, by the second and third terms and are positive 
since they are repulsive. We restrict ourselves to the physically relevant 
region $V \le U$ and set the scale of energies by using $t=1$.

Model~(\ref{eq:ebhmodel}) has been studied by a number of 
authors~\cite{sheshadri,tvr} in the case $V = 0$, i.e., in the absence of 
nearest-neighbor interactions; at zero temperature ($T = 0$) it has been 
shown to have a superfluid phase if the mean number of bosons per site 
$\rho$ is not an integer; however, for integer densities, it shows a 
superfluid (SF) to Mott insulator (MI) transition. This SF-MI transition is 
of the Kosterlitz-Thouless (KT) type~\cite{kt} in one 
dimension~\cite{fisher,batrouni,baltin,rvpai,kuhner,kashurnikov}.

In the limit $U \rightarrow \infty$ model~(\ref{eq:ebhmodel}) maps onto the 
spin-$\frac{1}{2}$ XXZ model if the mean number of bosons per site 
$\rho = 1/2$. 
Every site can now have only two possible states, namely, a state with no 
boson 
and another with one boson. We represent these as $\mid 0 \rangle$ and 
$\mid 1 \rangle$, respectively, and make the identification
$\mid 0 \rangle \, \rightarrow \, \mid \downarrow \rangle$ and
$\mid 1 \rangle \, \rightarrow \, \mid \uparrow \rangle$, where 
$\mid \downarrow \rangle$ and $\mid \uparrow \rangle$ are, respectively, 
spin-$\frac{1}{2}$ down and up states. Now, by using the transformations 
$a^{\dagger}_{i} \equiv S^+_{i}$, $a_{i} \equiv S^{-}_{i}$, and 
${\hat n}_{i} \equiv (S^{z}_i-\frac{1}{2})$, the model~(\ref{eq:ebhmodel}) 
maps 
onto the spin-$\frac{1}{2}$ XXZ model with the Hamiltonian
\begin{equation}
\label{eq:xxz}
      {\cal H}_{XXZ} = -2t \sum_{<i,j>} (S_i^x S_j^x+S_i^y S_j^y) + 
V \sum_{<i,j>} S_i^z S_j^z ,
\end{equation}
where we have suppressed constant terms. This model has been solved 
exactly~\cite{baxter} and shows a KT-type transition from XY to Ising 
ordering at $V=2t$. The bosonic analogs of XY and Ising phases are, 
respectively, SF and MDW phases.

If $\rho = 1$ and $t = 0$, it is easy to see that model~(\ref{eq:ebhmodel}) 
has a first-order, MI-MDW transition at $U = 2V$. Large values of $U$ favor
the MI phase whereas large values of $V$ favor the MDW phase.
 
Recently K\"{u}hner \textit{et al}~\cite{kuhner} studied 
model~(\ref{eq:ebhmodel}) in one dimension by the using a finite-size, 
density-matrix renormalization group~\cite{white} (FSDMRG) and showed that, 
for  $V=0.4$, it has a continuous SF-MDW transition for density 
$\rho=\frac{1}{2}$ and a continuous SF-MI transition for $\rho=1$.
Niyaz \textit{et al}~\cite{niyaz} have studied this model
in one dimension by a Monte-Carlo method and obtained its phase diagram
in the $(U,V)$ plane for $\rho = 1$. They obtain SF, MI and MDW phases in  
model~(\ref{eq:ebhmodel}) and continuous SF-MI, SF-MDW, and MI-MDW
transitions but conjecture that, at large $U$, the MI-MDW transition should
be first order. The study of Ref.~[\onlinecite{niyaz}] has obtained a phase 
diagram for 
model~(\ref{eq:ebhmodel}); however, they have not investigated the 
universality classes of the transitions in detail. 
We obtain the phase diagram  here for density  $\rho=1$ by using the 
FSDMRG method which, as we show 
below, gives very accurate results for the nature of ordering in the different 
phases and the types and universality classes of the transitions. We restrict 
ourselves to the case of integer density ($\rho = 1$) since we want to explore 
the competition between SF, MDW, and MI phases. We note in passing that, 
even for $V = 0$, the Bose-Hubbard model~(\ref{eq:ebhmodel}) cannot be solved 
exactly unlike its fermionic counterpart; and for the fermionic case too 
there has been renewed interest in the phase diagram of the extended Hubbard 
model~\cite{jackelmann}. 

\begin{figure}
\includegraphics{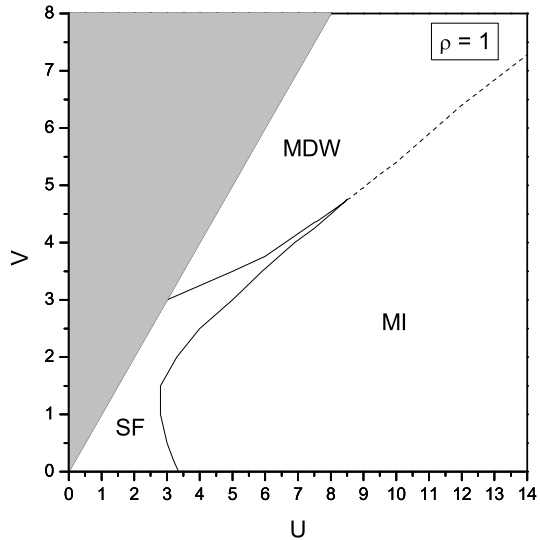}
\caption{\label{fig:phase_diagram}The FSDMRG phase diagram of the 
one-dimensional, extended Bose-Hubbard model for density $\rho=1$ showing 
SF, MI, and MDW phases. The two full lines indicate continuous transitions 
whereas the dashed line is a first-order boundary; these meet at a bicritical
point. We do not consider the shaded region $V>U$.}
\end{figure}

Before proceeding further we give a brief summary of our results. Our FSDMRG 
phase diagram for model~(\ref{eq:ebhmodel}), with $d = 1$ and $\rho=1$, 
is given in the $(U,V)$ plane of Fig.~(\ref{fig:phase_diagram}). It consists of
three phases; Superfluid (SF), Mott Insulator (MI) and Mass Density Wave (MDW).
For small values of the interactions $U$ and $V$, the SF phase dominates  
as is to be expected since the bosons interact weakly here. However, as the 
interaction strengths increase, either MI  or MDW phases get stabilized. 
The former dominates when $U$ is much larger than $V$, whereas the latter 
dominates if $U$ and $V$ are both large and comparable: A large, repulsive $V$ 
disfavors a phase with a uniform density of bosons on nearest-neighbor sites; 
instead, an MDW phase, with a periodic variation of the boson density, 
is stabilized. The lattice we consider is bipartite and has two sublattices 
A and B (say odd-numbered and even-numbered sites); the ground
state in the MDW phase is, therefore, doubly degenerate since the 
peaks in the mass-density wave can lie either on the A or the B 
sublattice. If the bosons are charged this MDW phase is a 
charge-density-wave (CDW) phase. By using the FSDMRG method we have determined 
the phase boundaries between these phases. The MI-SF phase boundary in 
Fig.~(\ref{fig:phase_diagram}) lies in the Kosterlitz-Thouless 
(KT) universality class, whereas the MDW-SF phase boundary has 
both KT and two-dimensional-Ising characters. For large values of $U$ and $V$,
the MI-MDW transition occurs directly and is of first-order [dashed line in 
Fig.~(\ref{fig:phase_diagram})]; as noted above, at $t = 0$, 
a first-order MI-MDW 
transition is obtained at $U = 2V$. Within the accuracy of our calculation, 
the MI-SF, MDW-SF, and MI-MDW phase boundaries meet at a bicritical point
at ($U ,V ) = (8.5 \pm 0.05 ,4.75 \pm 0.05)$. We have 
looked for, but not found, a supersolid phase with both SF and MDW order.
A very brief discussion of some of our preliminary results has been 
given in Ref.~[\onlinecite{rvpai_cnr}].

The remaining part of this paper is organized as follows. 
Section II contains the details of our Finite-size Density-Matrix 
Renormalization Group (FSDMRG) calculation. Section III contains our results.
We end with concluding remarks in Section IV.

\section{FSDMRG Calculations}

The Finite-Size Density-Matrix Renormalization Group (FSDMRG) method 
has proven to be very useful in studies of one-dimensional quantum 
systems~\cite{rvpai,white,rvpai_cnr}. To make this paper self-contained 
we summarize the salient points of this method. Open boundary conditions are 
preferred  for such calculations since the loss of accuracy with increasing 
system size is much less than in the case of periodic boundary conditions.
The conventional FSDMRG method consists of the following two steps:
\begin{enumerate}
\item The infinite-system density-matrix renormalization group method (DMRG),
where we start with a system with four sites, add two sites at each
step of the iteration, and continue till we obtain a system with the desired 
number $L$ of sites ( in most of our calculations we use $L \simeq 100$ but, 
in some representative cases, we have gone up to $L=200$).
\item The finite-system method in which the system size $L$ is held fixed, but
the energy of a target state is improved iteratively by a sweeping procedure,
described below, till convergence is obtained. 
\end{enumerate}
For a model like Eq.~(\ref{eq:ebhmodel}) we first construct the Hamiltonian 
matrix of the superblock configuration 
${\bf B}^{\ell}_1 \bullet \bullet ~{\bf B}^r_1$, where 
${\bf B}^{\ell}_1$ and ${\bf B}^r_1$ represent left- and right-block
Hamiltonians, respectively, and each one of the $\bullet$ represents a 
single-site Hamiltonian. In the first step of the DMRG iteration both 
${\bf B}^{\ell}_1$ and ${\bf B}^r_1$ also represent single sites, so, at this 
step, we have a four-site chain. We now diagonalize the Hamiltonian matrix of 
the superblock and obtain the energy and the eigenfunction of a  
\textit{target state}. In our study the target state is the ground 
state of the system of size $L$ with either $N=L$ or $N=L\pm 1$ bosons. 
The latter is required for obtaining the gap in the energy spectrum.  We now 
divide the superblock into two equal halves, the left and the right parts, 
which are treated, respectively, as the \textit{system} and the 
\textit{universe}. The density matrix for this \textit{system}, namely, 
${\bf B}^{\ell}_2 \equiv {\bf B}^{\ell}_1 ~\bullet$, is 
calculated from the \textit{target state}. If we write the 
\textit{target state} as 
$\mid \psi \rangle=\sum_{i,j} \psi_{i,j} \mid i\rangle \mid j \rangle$, where 
$\mid i\rangle$ and $\mid j \rangle$ are, respectively, the basis states of 
the \textit{system} and the \textit{universe}, then the density matrix for 
the \textit{system} has elements $\rho_{i,i'}= \sum_{j}\psi_{i,j}\psi_{i',j}$.
The eigenvalues of this density matrix measure the weight of each of its 
eigenstates in the \textit{target state}. The optimal states for describing 
the \textit{system} are the ones with the largest eigenvalues of the
associated density matrix. In this first step of the DMRG the
superblock, and hence the dimension of the density matrix, is small, so all the
states can be retained. However, in subsequent steps, when the sizes of the
superblocks and density matrices increase, only the most significant states
are retained, say the ones corresponding to the largest $M$ eigenvalues of
the density matrix (in our studies we choose $M = 128$). We then obtain the
effective Hamiltonian for the \textit{system} ${\bf B}^{\ell}_2$ in the basis 
of the significant eigenstates of the density matrix; this is used in turn as
the left block for the next DMRG iteration. In the same manner we obtain
the effective Hamiltonian for the right block, i.e.,  
${\bf B}^r_2 \equiv \bullet {\bf B}^r_1$. In the next step of the DMRG we 
construct the Hamiltonian matrix for the superblock  
${\bf B}^{\ell}_2 \bullet \bullet ~{\bf B}^r_2$, so the size of the system
increases from $L = 4$ to $L = 6$.  For a system of size $L$, we continue, 
as in the first step, by diagonalizing the Hamiltonian matrix for the 
configuration  
${\bf B}^{\ell}_{\frac{L}{2}-1} \bullet \bullet {\bf B}^r_{\frac{L}{2}-1}$ 
and setting 
${\bf B}^{\ell}_{\frac{L}{2}} \equiv {\bf B}^{\ell}_{\frac{L}{2}-1} \bullet$ 
and 
${\bf B}^r_{\frac{L}{2}}\equiv \bullet {\bf B}^r_{\frac{L}{2}-1}$
in the next step of the DMRG iteration. Thus at each step of the DMRG 
iteration the left and right blocks increase in length by one site and 
the total length $L$ of the chain increases by $2$.

In the infinite-system DMRG method outlined above the left- and right-block 
bases are not optimized in the following sense:  The DMRG estimate for the 
target-state energy, at the step when the length of the system 
is $L$, is not 
as close to the exact value of the target-state energy for this system 
size as it can be. It has been found that the FSDMRG method overcomes this 
problem~\cite{white}.  In this method we first use the infinite-system DMRG 
iterations to build up the system to a certain desired size $L$. The $L$-site 
superblock configuration is now given by 
${\bf B}^{\ell}_{\frac{L}{2}-1} \bullet ~ \bullet ~{\bf B}^r_{\frac{L}{2}-1}$. 
In the next step of the FSDMRG method, the superblock configuration 
${\bf B}^{\ell}_{\frac{L}{2}} ~ \bullet ~ \bullet ~{\bf B}^r_{\frac{L}{2}-2}$, 
which clearly keeps the system size fixed at $L$, is used. This step is called 
\textit{sweeping} in the right direction since it increases (decreases) 
the size of the left (right) block by one site. For this superblock the 
\textit{system} is ${\bf B}^{\ell}_{\frac{L}{2}} \bullet$, the 
\textit{universe} is $\bullet {\bf B}^r_{\frac{L}{2}-2}$, the associated 
density matrix can be found, and from its most significant states the new 
effective Hamiltonian for the left block, with $(\frac{L}{2} + 1)$ sites, is 
obtained. 
We sweep again, in this way, to obtain a left block with $(\frac{L}{2} + 2)$
sites and so on till the left block has $(L-3)$ sites and the right block
has 1 site so that, along with the two sites in between these blocks, the 
system still has size $L$; or, if a preassigned convergence criterion 
for the target-state energy is satisfied, this sweeping can be terminated  
earlier. Note that, in these sweeping steps, for the 
right block we need ${\bf B}^r_1$ to ${\bf B}^r_{L-3}$, which we have 
already obtained in  earlier steps of the infinite-system DMRG. Next we sweep 
leftward: the size of the left (right) block decreases  (increases) by one 
site 
at each step. Furthermore, in each of the right- and left- sweeping steps, the 
energy of the target state decreases systematically till it converges (we use a
six-figure convergence criterion in our calculations). 

\begin{figure}
\includegraphics{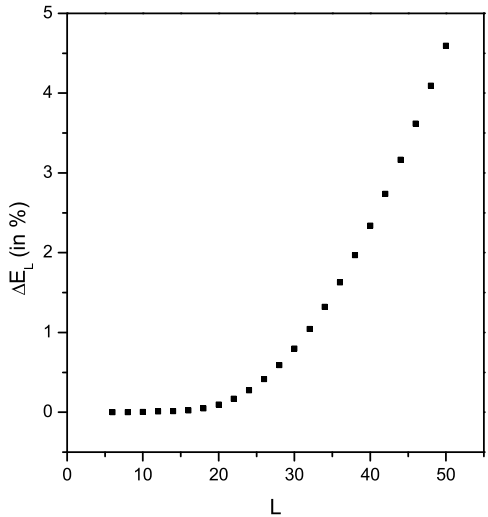}
\caption{\label{fig:fsdmrg} The percentage difference between the ground state
energies $\Delta E_L=(E^{DMRG}_L-E^{FSDMRG}_L)/E^{FSDMRG}_L$ 
obtained by DMRG and FSDMRG methods, for a representative 
point in the SF phase of Fig~(\ref{fig:phase_diagram}), plotted as a function
of $L$.}
\end{figure}

We use a slightly modified form of the FSDMRG method in which we sweep, as 
described above, at \textit{every step} of the DMRG scheme and not only in the 
one that corresponds to the largest value of $L$. This helps in obtaining
accurate $\beta$ functions which we use to obtain critical exponents 
(see below) at continuous transitions. Furthermore, since the superfluid 
phase in models such as Eq.~(\ref{eq:ebhmodel}), in $d = 1$ and at $T = 0$, 
is critical and has a correlation length that diverges with the system
size $L$, finite-size effects must be removed by using finite-size scaling
as we show below. For this purpose, the  energies and  
correlation functions, obtained from a DMRG calculation, should have converged 
properly for each system size $L$. It is important, therefore, that we use the
FSDMRG method as opposed to the infinite-system DMRG method especially in the
vicinities of continuous phase transitions. We find that convergence, to a 
specified accuracy for the target-state energy, is faster in the MI phase 
than in the SF phase. Figure~(\ref{fig:fsdmrg}) shows, at a representative 
point in the SF phase of Fig.~(\ref{fig:phase_diagram}), how the 
percentage disagreement 
between DMRG and FSDMRG ground-state energies  increases with $L$ in our 
calculations. 

Since the bases of left- and right-block Hamiltonians are truncated by 
neglecting the eigenstates of the density matrix corresponding to small 
eigenvalues, this leads to truncation errors. If we retain $M$ states, the 
density-matrix weight of the discarded states is
$P_M=\sum_{\alpha=1}^{M} (1-\omega_\alpha)$, where $\omega_{\alpha}$ are the
eigenvalues of density matrix. $P_M$ provides a convenient measure of the 
truncation 
errors. We find that these errors depend on the order-parameter correlation 
length in a phase. For a fixed $M$, we find very small truncation errors in 
the MI and MDW phases; these grow as the MI-SF and MDW-SF transitions are 
approached; and the truncation errors are largest in the SF phase. In our 
calculations we choose $M$ such that the truncation error is always less than 
$5\times 10^{-6}$; we find that $M = 128$ suffices.

The number of possible states per site in the Bose-Hubbard model is 
infinite since there can be any number of bosons on a site. In a practical 
DMRG calculation we must restrict the number $n_{max}$ of states or bosons 
allowed per site. The smaller the interaction parameters $U$ and $V$, the 
larger must $n_{max}$ be. 
As in earlier calculations~\cite{rvpai,kuhner,rvpai_cnr} on
related models, we find that  $n_{max} = 4$ is sufficient for the values of 
$U$ and $V$ considered here; we have checked in representative cases that 
our results do not change significantly if $n_{max} = 5$. 

In summary, then, our FSDMRG procedure gives us the energy $E_L(N)$ for the
ground state of model~(\ref{eq:ebhmodel}) and the associated eigenstate
$\mid \psi_{0LN} \rangle$. Given these we can calculate the energy gaps, order
parameters, and correlation functions that characterize all the phases of this
model and thence the phase diagram. We discuss this in Section III.

\section{Results}

The single-particle energy gap $G_L$ for a system of size $L$, the 
order parameter for the MDW phase, and the correlation functions 
that characterize SF and MDW phases in model~(\ref{eq:ebhmodel}) can 
be defined in a straightforward manner in terms of the energies and 
wavefunctions mentioned in the previous Section. The energy gap is  
\begin{equation}
G_L=E_L(N+1)+E_L(N-1)-2 E_L(N) ,
\label{eq:gap}
\end{equation}
where $E_L(N)$ is the ground-state energy for a system of size $L$ with $N$ 
bosons; since we are interested in studying the case $\rho = 1$, we 
increase the number of bosons by 2 at every DMRG step in which 2 sites
are added to the system (Sec. II) so that $\rho = N/L = 1$. We expect, and 
show explicitly below, that this gap is positive in both MI and MDW phases, 
which are incompressible insulators, but it vanishes in the SF phase, which is 
compressible.

The correlation function that characterizes the SF phase is 
\begin{equation}
 \Gamma_L^{SF}(r)\equiv \langle
\psi_{0LN}| a_0^\dagger a_r |\psi_{0LN}\rangle ,
\label{eq:gammasf}
\end{equation}
where $|\psi_{0LN}\rangle$ is the ground-state wavefunction of the  
system with size $L$ and $N$ bosons. The associated correlation length can 
be obtained from the second moment of this correlation function, namely,
\begin{equation}
\xi^{SF}_L \equiv \left[\frac{\sum_r r^2 \Gamma_L^{SF}(r)}
{\sum_r \Gamma_L^{SF}(r)}\right]^{1/2} .
\label{eq:corrlensf}
\end{equation}
Note that $\xi^{SF}_L$ is the correlation length for SF ordering in a 
system of size $L$; it remains finite so long as $L < \infty$.

The MDW phase can be differentiated from the MI phase by using the 
order parameter for the MDW phase 
\begin{equation}
  \label{eq:o_dw}
  M_{MDW}=\frac{1}{L}\sum_i (-1)^{i}\langle\psi_{0LN}| 
(\hat{n}_i - \rho)|\psi_{0LN}
\rangle,
\end{equation}
and the associated correlation function 
\begin{equation}
\Gamma_L^{MDW}(r)\equiv  \langle\psi_{0LN}|
(\hat{n}_0 - \rho)(\hat{n}_{r} - \rho)|\psi_{0LN}\rangle;
\label{eq:corrmdw}
\end{equation}
the correlation length for MDW ordering $\xi^{MDW}_L$ can be defined as in 
Eq.~(\ref{eq:corrlensf}) but with $\Gamma_L^{MDW}$ instead of $\Gamma_L^{SF}$.

It is useful first to discuss the case $V=0$. Here we reproduce the 
well-understood SF-MI transition. In $d = 1$ the appearance of the SF 
phase is signaled by the divergence of the correlation 
length $\xi^{SF}_{L\to\infty}$. For a finite system $\xi^{SF}_{L}$ is 
finite and we must extrapolate to the $L\rightarrow \infty$ limit,
which is best done by using finite-size scaling. In the critical region 
the correlation length
\begin{figure}
\includegraphics{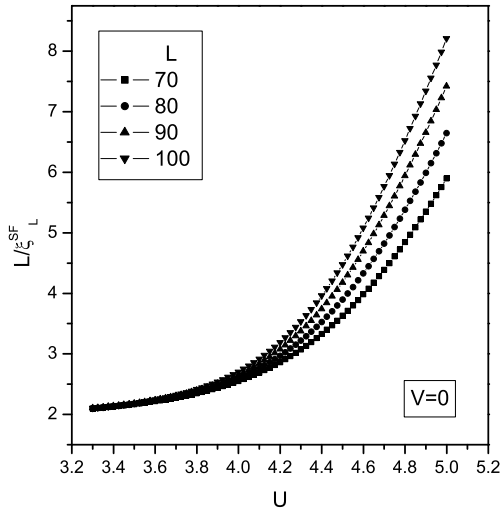}
\caption{$L/\xi_L^{SF}$ plotted as a function of $U$ for 
    different system sizes and $V=0$ in model (\ref{eq:ebhmodel}).
    The coalescence of different curve for $U\simeq3.4$ shows a 
Kosterlitz-Thouless-type SF-MI transition.}
  \label{fig:v00_xi}
\end{figure}
\begin{equation}
  \label{eq:scaling-corr}
  [\xi^{SF}_L]^{-1} \approx L^{-1}f(L/\xi^{SF}_\infty),
\end{equation}
where the scaling function $f(x) \sim x , \, x \to 0$. Thus  plots of 
$L/\xi^{SF}_L$ versus $U$, for different system sizes $L$, consist of curves 
that intersect at the critical point at which the correlation length for
$L=\infty$ diverges. Such plots are given in Fig.~(\ref{fig:v00_xi}) 
for $V=0$. Curves for different values of $L$ coalesce for 
$U \le U_c \simeq 3.4$ 
indicating the existence of a critical SF phase, with a diverging correlation
length, i.e., a power-law decay of correlations, for all $0 \le U \le U_c$. 
The single-particle gap $G_L$ scales as the inverse of this correlation length 
and is, therefore, zero for $L = \infty$ for $0\le U \le U_c$. 
For $U > U_c$ we have an MI phase with a finite correlation length and a 
nonzero gap. 
Figure~(\ref{fig:v00_xi}) also suggests that the SF-MI transition here is of 
the KT type. We can quantify this by calculating the $\beta$ function at this
transition via the Roomany-Wyld (RW) approximants~\cite{rvpai,roomany,barber}
\begin{equation}
  \label{eq:roomany}
  \beta_{LL'}=\frac{1-\ln(\xi_{L}^{SF}/\xi_{L'}^{SF})/\ln(L/L')}
  {(\xi_L^{SF} \xi_{L'}^{SF}/\xi'_L {\xi'}_{L'})^{1/2}},
\end{equation}
where $L$ and $L'$ denote two system sizes and $\xi'_L \equiv d\xi_L^{SF}/dU$.
For a KT transition the correlation length 
$\xi_{\rm{KT}} \sim \exp(c/(U-U_c)^\sigma)$ and  
$\beta \sim (U-U_c)^{1+\sigma}$, with $\sigma=1/2$.
\begin{figure}
\includegraphics{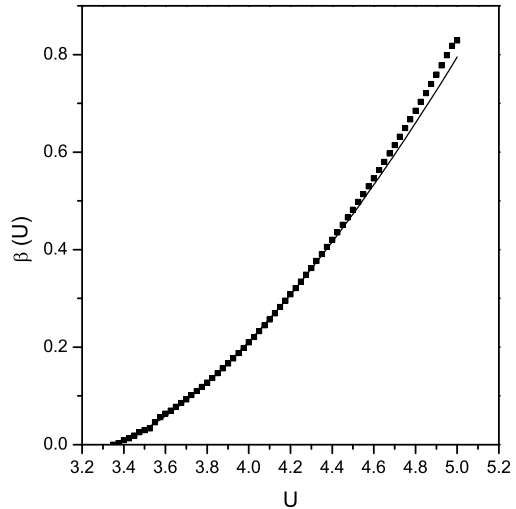}
  \caption{The $\beta$ function for the MI-SF transition for $V=0$ obtained 
by using RW approximants [Eq.~(\ref{eq:roomany})] with $L = 98$ and $L' = 100$.
The full line is a fit to the form $\beta(U)=c (U - U_c)^{(1+\sigma)}$; we get
$c=0.37 \pm 0.01$, $U_c=3.35 \pm 0.02$ and $\sigma=0.48 \pm 0.05$.}
  \label{fig:v00_beta}
\end{figure}
In Fig.~(\ref{fig:v00_beta}) we show the $\beta$ function for the SF-MI 
transition for 
$V=0$. To obtain this $\beta$ function we use  $L = 98$ and $L' = 100$ in 
Eq.~(\ref{eq:roomany}). Our fit to the data of Fig.~(\ref{fig:v00_beta}) yields
$U_c=3.35 \pm 0.02 $ and $\sigma = 0.48\pm 0.05$ which are consistent
with the values reported earlier~\cite{rvpai,kuhner}. If we fit our data (here
and below) over a fixed region of $(U-U_c)/U_c$, then our nonlinear 
least-squares programme yields smaller errors. The conservative error bars we
quote here reflect the range over which our fitted values for $U_c$, $\sigma$, 
etc., vary when we change the region of $(U-U_c)/U_c$ over which we fit our
data, namely, $0.001 < |U-U_c|/U_c < 0.2 $  or $0.001 < |U-U_c|/U_c < 0.35 $.  

\begin{figure}
\includegraphics{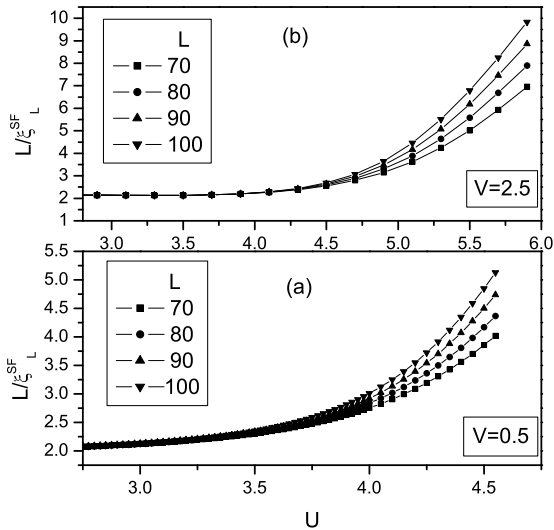}
\caption{$L/\xi_L^{SF}$  plotted as a function of $U$ for different system 
sizes and for  (a) $V=0.5$ and (b) $V=2.5$ in model (\ref{eq:ebhmodel}).
The coalescence of different curve for $U<3.0$ in (a) and $U < 4.2$ in (b) 
show KT-type SF-MI transitions.}
\label{fig:v05_xi}
\end{figure}
\begin{figure}
\includegraphics{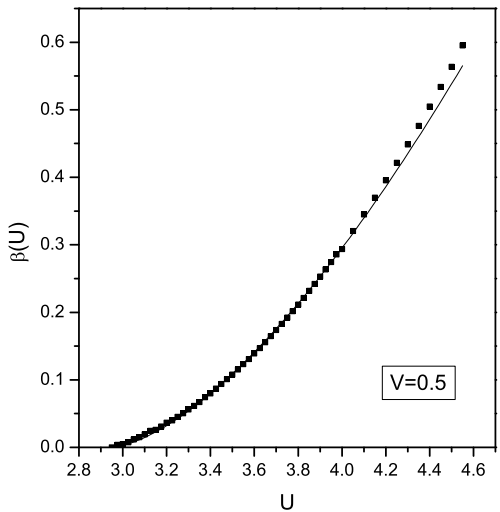}
\caption{The $\beta$ function of the MI-SF transition for $V=0.5$,
obtained by using the RW approximants with $L = 98$ and $L' = 100$ as
in Fig.~(\ref{fig:v00_beta}).
The full line is a fit to the form $\beta(U)=c (U - U_c)^{(1+\sigma)}$; we get
$c=0.27 \pm 0.01$, $U_c=2.95 \pm 0.02$ and $\sigma=0.53 \pm 0.05$.}
 \label{fig:v05_beta}
\end{figure}

Initially the nearest-neighbor interaction $V$ suppresses the SF phase relative
to the MI phase, but at larger values of $V$ this trend is reversed leading
to a reentrant SF [Fig.~(\ref{fig:phase_diagram})]. Figure~(\ref{fig:v05_xi}a) 
shows a 
plot of $L/\xi_L^{SF}$ versus $U$ for $V=0.5$; by comparing this with 
Fig.~(\ref{fig:v00_xi}) we see that $U_c(V=0.5) < U_c(V=0)$. Curves for 
different values of $L$ coalesce for $U \le U_c(V=0.5) \simeq 3.0$ 
indicating a power-law 
SF phase, as for $V = 0$, and a KT-type MI-SF transition. Again we use RW 
approximants to obtain the $\beta$ function shown in Fig.~(\ref{fig:v05_beta})
 and our numerical fit yields $U_c=2.95 \pm 0.02$ and $\sigma = 0.53 \pm 0.05$.
Eventually the MI-SF phase boundary in Fig.~(\ref{fig:phase_diagram}) turns 
back; a representative plot of $L/\xi_L^{SF}$ versus $L$, for $V = 2.5$, 
illustrates this
[Fig.~(\ref{fig:v05_xi}b)]; we find $U_c(V=2.5) \simeq 4.2$. This reentrance 
of the SF phase, with increasing $V$, was not resolved by the study 
of Ref.~[\onlinecite{niyaz}].
 
\begin{figure}
\includegraphics{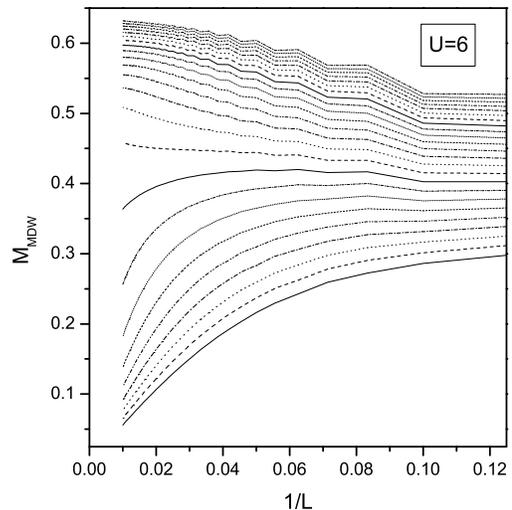}
\caption{The MDW order parameter $M_{MDW}$ versus $1/L$ for $U=6$ and  
different values of $V$ starting from 
  $3.0$ below  to $5.3$ above in steps of $0.1$.}
\label{fig:u60mdw_L}
\end{figure}

\begin{figure}
\includegraphics{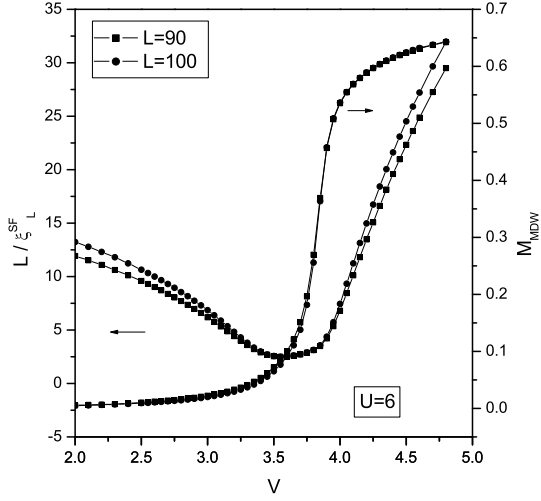}
\caption{Plots of $L/\xi^{SF}_L$ and $M_{MDW}$ versus $V$ for  
$U = 6$ and $L=90$ and $100$. The coalescence of different curves of 
$L/\xi^{SF}_L$, for 
$3.5< V < 3.9$, shows an SF phase sandwiched between MI and MDW phases;
for $3.9 \lesssim V$ we obtain the MDW phase with $M_{MDW} > 0$.}
\label{fig:u60xi_mdw}
\end{figure}

\begin{figure}
\includegraphics{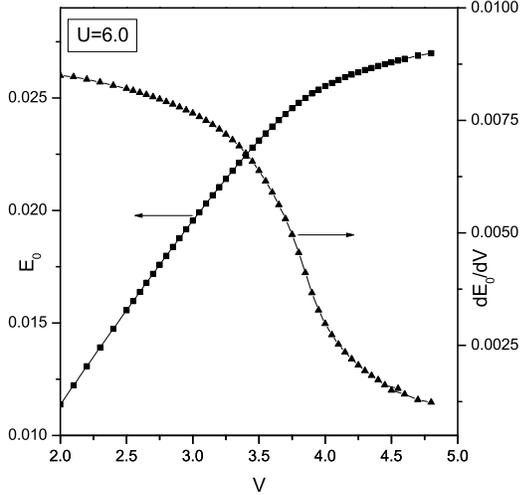}
\caption{Plots of the ground-state energy $E_0$ and its derivative $dE_0/dV$ 
versus $V$ for $U = 6$. There is no visible discontinuity in $dE_0/dV$ 
so both MI-SF and SF-MDW transitions are continuous.}
\label{fig:u60_e0}
\end{figure}

\begin{figure}
\includegraphics{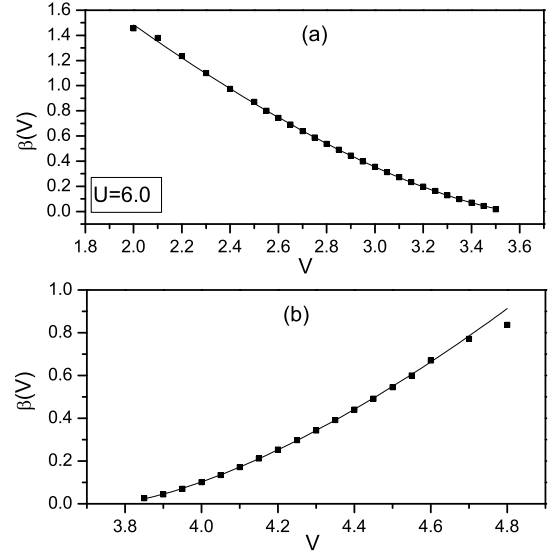}
\caption{The $\beta$ functions for MI-SF (a) and MDW-SF (b) transition for 
$U=6$,
obtained by using  RW approximants with $L = 98$ and $L' = 100$.
The full line is a fit to the form $\beta(U)=c (U - U_c)^{(1+\sigma)}$.
For the MI-SF transition (a) we get 
$c=0.76 \pm 0.01$, $U_c=3.59 \pm 0.05$ and $\sigma=0.47 \pm 0.05$ and
for the MDW-SF transition (b) we get 
$c=0.86 \pm 0.01$, $U_c=3.78 \pm 0.05$ and $\sigma=0.49 \pm 0.05$.
}
\label{fig:u60_beta}
\end{figure}

\begin{figure}
\includegraphics{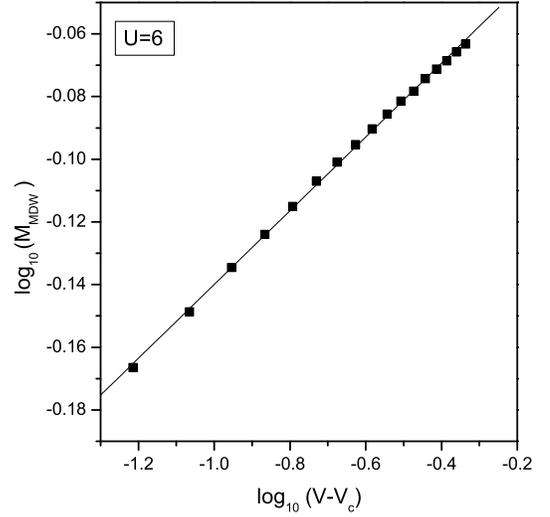}
\caption{A log-log plot (base 10) of the MDW order parameter $M_{MDW}$ versus
$V-V_c$ for $U=6$. The straight line is a fit to the form 
$M_{MDW} \sim (V-V_c)^{\beta_{MDW}}$, $V\downarrow V_c$. Our fit yields 
$V_c=3.87\pm0.05$ and $\beta_{MDW}=0.12\pm0.01$.}
\label{fig:u60_mdw}
\end{figure}

For sufficiently large values of $V$ we can have an MDW phase and an SF-MDW
transition, at small values of $U$, and an MI-MDW transition at large values 
of $U$ as shown in the phase diagram of Fig.~(\ref{fig:phase_diagram}). 
In Fig.~(\ref{fig:u60mdw_L}) we 
plot the MDW order parameter $M_{MDW}$ as a function of $1/L$ for $U = 6$ and 
values of $V$ 
ranging from $V=3.0$ to $5.3$ in steps of $0.1$. We see that $M_{MDW}$ goes to 
zero for $V < V_c(U=6) \simeq 3.9$ whereas it is nonzero for higher values of 
$V$.
To determine the universality classes of the MI-SF and SF-MDW transitions
we plot in Fig.~\ref{fig:u60xi_mdw}  both the order parameter $M_{MDW}$ 
and $L/\xi^{SF}_L$ as functions of $V$ for $U=6$. The different curves for  
$L/\xi^{SF}_L$ coalesce in the region $3.5 \lesssim U \lesssim 3.9$ indicating 
an SF phase sandwiched between MI and MDW phases. Both SF-MI and SF-MDW 
transitions are continuous. To confirm this we have also obtained plots of the
ground-state energy $E_0 \equiv \lim_{L\to\infty} E_L(N)$ as a function of $V$ 
for fixed $U$. In Fig.~(\ref{fig:u60_e0}) we plot $E_0$ and $dE_0/dV$ 
versus $V$; this 
plot shows no discontinuity at the SF-MI and SF-MDW transitions 
(as it does at the first-order MI-MDW transition discussed below for $U=12$).
To determine the universality classes of these SF-MI and SF-MDW transitions we 
have obtained $\beta$ functions, via RW approximants for $\xi^{SF}_L$, in  
Fig.~(\ref{fig:u60_beta}) for 
$U = 6.0$. For the MI-SF transition, we get $V_c = 3.59 \pm 0.05$ and $\sigma
= 0.47 \pm 0.05$; and, for the SF-MDW transition, $V_c=3.78 \pm 0.05$ and 
$\sigma = 0.49 \pm 0.05 $. Thus both of these transitions are of the 
KT type. However, in addition, the MDW-SF transition also
has an Ising character (two-dimensional) since the MDW phase has a doubly 
degenerate ground state as mentioned above.
To extract such Ising-model exponents, we use the form  
$M_{MDW}\sim (V-V_c)^{\beta_{MDW}}$ as $V\downarrow V_c$, where $\beta_{MDW}$
 is the MDW order-parameter exponent. For $U=6$ our fit yield 
$V_c=3.87\pm 0.05$ and   
$\beta_{MDW}=0.12 \pm 0.01$ (Fig.~\ref{fig:u60_mdw}) 
in good agreement with the two-dimensional, Ising order parameter exponent.
Note that the value of $V_c$ obtained from this fit for $M_{MDW}$ is within 
error bars of that obtained from the $\beta$ function for the SF-MDW 
transition. Thus, within our calculation we cannot resolve a supersolid phase
which has long-range SF correlations and MDW ordering.

\begin{figure}
\includegraphics{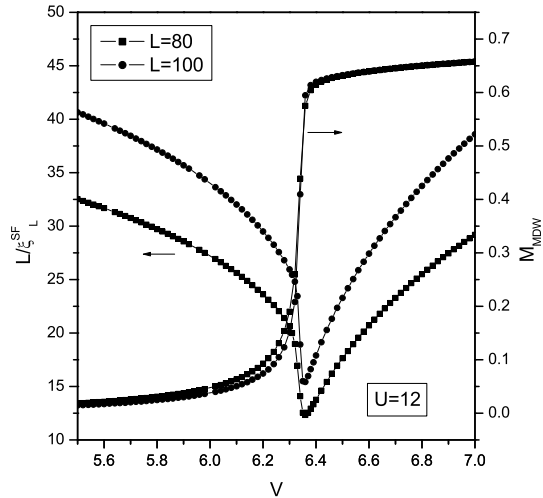}
\caption{Plots of $L/\xi_L^{SF}$ and $M_{MDW}$ versus $V$ for $U = 12$ and 
different 
system sizes $L$. Note that the curves for $L/\xi_L$ do not meet at any point;
however, $M_{MDW}$ jumps at the first-order, MI-MDW transition at
$V \simeq 6.3$.}
\label{fig:u12_lbyxi_mdw}
\end{figure}

\begin{figure}
\includegraphics{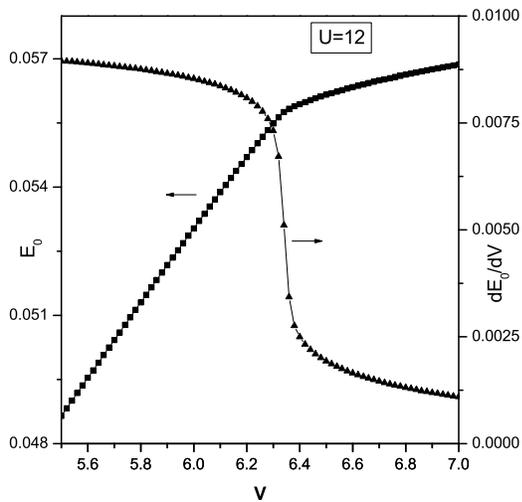}
\caption{The ground-state energy $E_0$ and its derivative $dE_0/dV$ versus 
$V$ for $U=12.0$; the jump in $dE_0/dV$ at $V \simeq 6.3$ shows that the 
MI-MDW transition is first order.}
\label{fig:u12_e0}
\end{figure}

From the phase diagram of Fig.~(\ref{fig:phase_diagram}) we see that, for 
sufficiently large values of $U$, there is no SF phase and only a direct, 
first-order MI-MDW transition. This direct transition shows up clearly in 
Fig.~\ref{fig:u12_lbyxi_mdw} where we plot $L/\xi_L^{SF}$ and $M_{MDW}$ 
versus $V$ for $U=12$: Curves of $L/\xi_L^{SF}$, for different values of $L$, 
do not merge 
at any point, so we can conclude that no power-law SF phase intervenes
between MI and MDW phases. Furthermore, the sharp jump in $M_{MDW}$ at
$V \simeq 6.3$ indicates that we have a first-order MI-MDW transition. This 
is corroborated by the plots of the ground-state energy $E_0$ and its 
derivative $dE/dV$ given in Fig.~\ref{fig:u12_e0} for $U = 12$; the
discontinuity of $dE/dV$ at $V \simeq 6.3$ indicates the first-order nature 
of the transition.

\begin{figure}
\includegraphics{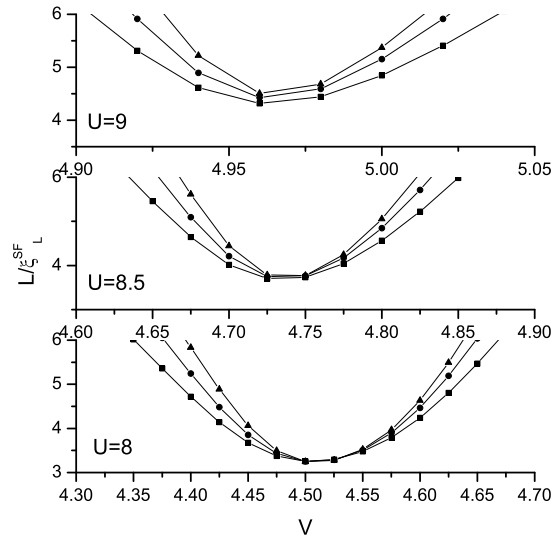}
\caption{Plots of $L/\xi_L^{SF}$ versus $V$ for different values of $U$ and
different system sizes $L (80,90,100)$. The region over which curves for 
different values of $L$ coalesce decreases as we go from $U=8$ to $U=8.5$; and
there is no coalescence for $U=9$. This  shows how the  SF phase shrinks 
as we approach the bicritical point.}
\label{fig:lbyxi_us}
\end{figure}

\begin{figure}
\includegraphics{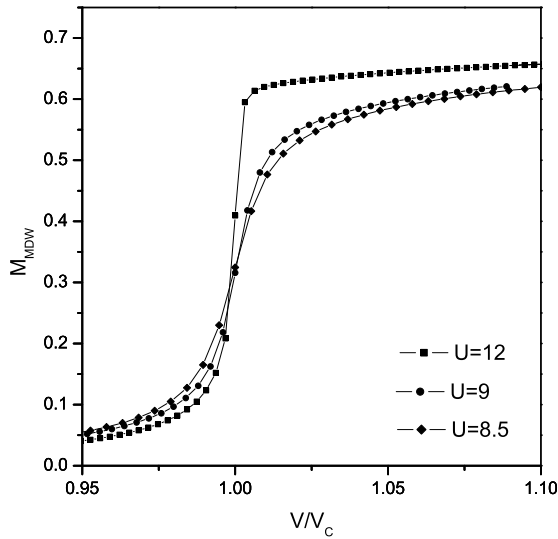}
\caption{The MDW order parameter $M_{MDW}$ plotted versus $V/V_c$ 
for different values of $U$ across the MI-MDW first-order transition. 
The jump in $M_{MDW}$
at the first-order phase boundary decreases as we approach the 
bicritical point at  ($U ,V ) = (8.5 \pm 0.05 ,4.75 \pm 0.05)$.}
\label{fig:m_mdw_us}
\end{figure}

The only feature of the phase diagram of Fig.~(\ref{fig:phase_diagram}) that 
remains
unexplored now is the region in which the continuous MI-SF and SF-MDW phase
boundaries meet the first-order MI-MDW boundary. The simplest topology 
possible here is that these meet at one \textit{bicritical} point. Our data 
are not inconsistent with such a topology. Figure~(\ref{fig:lbyxi_us}) shows,
via plots of $L/\xi^{SF}_L$ versus $V$, for different values of $L$ and $U$, 
how the extent of the SF phase (the region over which the curves of
$L/\xi^{SF}_L$ coalesce) shrinks as we approach the bicritical point. Similarly
Fig.~(\ref{fig:m_mdw_us}) shows how the jump in $M_{MDW}$ across the 
MI-MDW first-order transition decreases as we approach the bicritical point.
To confirm that this is indeed the topology of the phase diagram in the 
region where the MI-SF, SF-MDW, and MI-MDW phase boundaries meet, we 
must obtain the critical exponents in the vicinity of the bicritical point.
This is beyond the accuracy of our calculation at the moment. Nor can we rule
out completely more complicated topologies of phase diagrams in which
very closely spaced tricrital points and critical endpoints are used to link 
the three phase boundaries we have studied above. 

\section{Conclusions}

	We have carried out an extensive study of the one-dimensional, 
extended Bose-Hubbard model~(\ref{eq:ebhmodel}) by using the FSDMRG method.
Our study yields ground-state energies, single-particle gaps, the 
MDW order parameter, and SF correlation functions and correlation lengths. 
By studying these
we obtain an accurate phase diagram [Fig.~(\ref{fig:phase_diagram})] for this 
model.
This shows continuous MI-SF and SF-MDW transitions meeting the first-order
MI-MDW boundary at a bicritical point.

	We have looked for, but not found, a supersolid (SS) phase which, 
in the context of the lattice model we study here, would exhibit power-law
superfluid correlations, as in the SF phase, and a nonzero order parameter
$M_{MDW}$, as in the MDW phase. It is likely~\cite{frey} that further
than nearest-neighbor interactions will be required to stabilize the SS phase
as we will explore elsewhere.

	We hope our detailed study of model~(\ref{eq:ebhmodel}) will stimulate
experimental studies. A recent experimental study~\cite{stoferle} has shown 
that it is possible, by a suitable choice of 
confining potentials in optical lattices, to obtain a physical realization of 
the one-dimensional Bose-Hubbard
models. It would be interesting to see how nearest-neighbor interactions, 
like $V$, can be obtained in such lattices. If this can be done, the rich 
phase diagram of Fig.~(\ref{fig:phase_diagram}) can be explored 
experimentally.

\begin{acknowledgments}
We would like to thank H.R. Krishnamurthy for discussions.
One of us (RVP) thanks the Jawaharlal Nehru Centre for Advanced Scientific
Research and the Department of Physics, Indian Institute of Science, Bangalore 
for hospitality during the time when a part of this paper was written. This 
work was supported by DST, India 
(Grants No. SP/S2/M-60/98 and SP/I2/PF-01/2000) and UGC, India.
\end{acknowledgments}


\begin{thebibliography}{11}
\bibitem{ma}M. Ma, B. I. Halperin, and P. A. Lee,
  Phys. Rev. B \textbf{34} 3136 (1989);
  P. Nisamaneephong, L. Zhang, and M. Ma, 
  Phys. Rev. Lett. \textbf{71} 3830 (1993).

\bibitem{fisher}
  M. P. A. Fisher, P. B. Weichman, G. Grinstein and D. S. Fisher, 
  Phys. Rev. B \textbf{40} 546 (1989).
  
\bibitem{krauth}
  W. Krauth, M. Caffarel, and J. P. Bouchaud, 
  Phys. Rev. B \textbf{45} 3137 (1992).

\bibitem{sheshadri}
  K. Sheshadri, H. R. Krishnamurthy, R. Pandit, and T. V. Ramakrishnan,
  Europhys. Lett. \textbf{22} 257 (1993); 
  Phys. Rev. Lett. \textbf{75} 4075 (1995).

\bibitem{tvr} 
T. V. Ramakrishnan,
Ordering Disorder: Prospect and Retrospect in 
  Condensed Matter Physics,
eds. V. Srivastava, A. K. Bhatnagar and D. G. Naugle
AIP Conference Proceedings \textbf{286} 
38 (1994) and references therein.

\bibitem{amico}
  L. Amico and V. Penna, 
  Phys. Rev. Lett. \textbf{80} 2189 (1998).

\bibitem{rvpai}
  R. V. Pai, R. Pandit, H. R. Krishnamurthy and S. Ramasesha,
  Phys. Rev. Lett. \textbf{76} 2937 (1996).

\bibitem{baltin}
  R. Baltin  and K. H. Wagenblast, 
  Europhys. Lett. \textbf{39} 7 (1997).

\bibitem{kuhner}
  T. D. K\"{u}hner  and H. Monien, 
  Phys. Rev. B \textbf{58} R14741 (1998),
  T. D. K\"{u}hner, S. R. White  and H. Monien H, 
  Phys. Rev. B \textbf{61} 12474 (2000).

\bibitem{kashurnikov}
  V. A. Kashurnikov and B. V. Svistunov, 
  Phys. Rev. B. \textbf{53} 11776 (1996).

\bibitem{batrouni}
  G. G. Batrouni, R. T. Scalettar, G. T. Zimanyi and A. P. Kampf, 
  Phys. Rev. Lett. \textbf{74} 2527 (1995);
  P. Niyaz, R. T. Scalettar, C. Y. Fong and G. G. Batrouni,
  Phys. Rev. B. \textbf{44} 7143 (1991).

\bibitem{nandini}
  W. Krauth, N. Trivedi, and D. Ceperley, 
  Phys. Rev. Lett. \textbf{67} 2307 (1991);
  N. Trivedi and M. Makivic, 
  Phys. Rev. Lett. \textbf{74} 1039 (1995).

\bibitem{wallin}
  M. Wallin, E. S. Sorensen, S. M. Girvin, and A. P. Young, 
  Phys. Rev. B \textbf{49} 12115 (1994).

\bibitem{zhang}
  S. Zhang, N. Kawashima, J. Carlson, and J. E. Gubernatis, 
  Phys. Rev. Lett. \textbf{74} 1500 (1995).

\bibitem{chan}
  M. H. W. Chan, K. I. Blum, S. Q. Murphy, G. K. S. Wong, and J. D. Reppy, 
  Phys. Rev. Lett. \textbf{61} 1950 (1988).

  
\bibitem{chow}
  E. Chow, P. Delsing and D. B. Haviland, 
  Phys. Rev. Lett \textbf{81} 204 (1998).
  
\bibitem{Mooji}
  A. van Oudenaarden, and J. E. Mooij, 
  Phys.  Rev. Lett. \textbf{76} 4947 (1996);
  A. van Oudenaarden, B. van Leeuwen, M. P. M. Robbens and J. E. Mooij, 
  Phys.  Rev. B \textbf{57} 11684 (1998).

\bibitem{goldman}
  D. B. Haviland, Y. Liu, and A. M. Goldman,
  Phys. Rev. Lett. \textbf{62} 2180 (1989).

\bibitem{nelson}
  D. R. Nelson and V. M. Vinokur, 
  Phys. Rev. B \textbf{48} 13060 (1993).

\bibitem{greiner}
  M. Greiner, O. Mandel, T. Esslinger, T. W. H\"{a}nsch and L. Bloch
  Nature (London) \textbf{415} 39 (2002).
  
\bibitem{jaksch}
  D. Jaksch, C. Bruder, J. I. Cirac, C. W. Gardiner and P. Zoller,
  Phys.  Rev. Lett. \textbf{81} 3108 (1998).

\bibitem{kim} 
E. Kim and M. H. W. Chan, 
Nature (London)  \textbf{427} 225 (2004).

\bibitem{kt}
J. M. Kosterlitz and D. J. Thouless, 
J. Phys. C \textbf{6}, 1181 (1973).

\bibitem{baxter}
R. J. Baxter,
Exactly Solved Models in Statistical Mechanics,
(Academic-Press, New York, 1982).

\bibitem{white}
S. R. White,
Phys. Rev. Lett. \textbf{69} 2863 (1992);
Phys. Rev. B. \textbf{48} 10345 (1993).


\bibitem{niyaz}
P. Niyaz, R. T. Scalettar, C. Y. Fong and G. G. Batrouni,
Phys. Rev. B. \textbf{50} 362 (1994).

\bibitem{jackelmann} 
E. Jeckelmann, 
Phys. Rev. Lett. \textbf{89} 236401 (2002).

\bibitem{rvpai_cnr} 
R. V. Pai and R. Pandit, 
Proc. Indian Academy of Sciences 
(Chemical Sciences)  \textbf{115} 721 (2003).

\bibitem{roomany}
H. H. Roomany and H. W. Wyld,
Phys. Rev. D \textbf{21} 3341 (1980).

\bibitem{barber}
M. N. Barber,
Phase Transition and Critical Phenomena,
eds. C. Domb C. and J. L. Lebowitz,
\textbf{8} (Academic, New York)
145 (1990).

\bibitem{frey} 
E. Frey and L. Balents,
Phys. Rev. B \textbf{55} 1050 (1997).

\bibitem{stoferle}

T. St\"{o}ferle, H. Moritz, C. Schori, M. K\"{o}hl and T. Esslinger,
Phys. Rev. Lett. \textbf{92} 130403 (2004).
\end{thebibliography}
\end{document}